\documentstyle[psfig,amsmath]{mn}

\title[On the evolution of young radio-loud AGN]
{On the evolution of young radio-loud AGN}

\author[I. Snellen et al.]{I.A.G. Snellen$^{1,2}$, R.T. Schilizzi$^{2,3}$, G.K. Miley$^{2}$, A.G. de Bruyn$^{4,5}$,\cr M.N. Bremer$^{2,6,7}$, H.J.A. R\"ottgering$^{2}$\\ 
$^{1}$Institute of Astronomy, Madingley Road, Cambridge CB3 0HA, United 
Kingdom\\
$^{2}$Leiden Observatory, P.O. Box 9513, 2300 RA, Leiden, The Netherlands \\
$^{3}$Joint Institute for VLBI in Europe, Postbus 2, 7990 AA, Dwingeloo, 
The Netherlands\\
$^{4}$Netherlands Foundation for Research in Astronomy, Postbus 2, 7990 AA, 
Dwingeloo, The Netherlands\\
$^{5}$Kapteyn Institute, Postbus 800, 9700 AV, Groningen, The Netherlands\\
$^{6}$Institut d'Astrophysique de Paris, 98bis Boulevard Arago, 75014 
         Paris, France\\
$^{7}$ Department of Physics, Bristol University, H H Wills Physics Laboratory,  Tyndall Avenue, Bristol, BS8 1TL, United Kingdom}

\date{}
\begin{document}
\maketitle
\begin{abstract}
This paper describes an investigation of the early evolution of extra-galactic 
radio sources using samples of faint and bright Gigahertz Peaked Spectrum
(GPS) and Compact Steep Spectrum (CSS) radio galaxies.
Correlations found between their peak frequency, peak flux density and 
angular size provide strong evidence that synchrotron self absorption is the 
cause of the spectral turnovers, and indicate that young radio sources 
evolve in a self-similar way. In addition, the data seem to suggest that 
the sources are in equipartition while they evolve.
If GPS sources evolve to large size radio sources, their redshift dependent 
birth-functions should be the same. Therefore, since the
lifetimes of radio sources are thought to be short compared to the Hubble 
time, the observed difference in redshift distribution between GPS and 
large size sources must be due to a difference in slope of their luminosity 
functions.
We argue that this slope is strongly affected by the luminosity evolution of 
the individual sources. A scenario for the luminosity evolution 
is proposed in which 
GPS sources increase in luminosity and large scale radio sources decrease
in luminosity with time. 
This evolution scenario is expected for a ram-pressure confined 
radio source in a surrounding medium with a King profile density.
In the inner parts of the King profile, the density of the medium is constant
and the radio source builds up its
luminosity, but after it grows large enough the density of the surrounding 
medium declines and the luminosity of the radio source decreases.
A comparison of the local luminosity function (LLF) of GPS galaxies with that
of extended sources is a good test for this evolution scenario.
Unfortunately, only a handful of GPS sources are known at low redshift, 
and an LLF can only be derived, assuming that their  
cosmological number density evolution is similar to that 
of steep spectrum sources. The LLF derived in this way is shown to be
in good agreement with the proposed evolution scenario. However, the
uncertainties are large, and larger, homogeneously selected samples of 
GPS sources are needed.

\end{abstract}

\section{Introduction}
\subsection{Gigahertz Peaked Spectrum and Compact Symmetric Objects.}

An important element in the investigation of the evolution of
extra-galactic radio sources is the study of young counterparts of
`old' FRI/FRII extended objects. Two classes of compact radio source
can be found in the literature as most likely representatives of this
early evolutionary stage: I) Gigahertz Peaked Spectrum (GPS) sources,
which are characterised by a convex-shaped radio spectrum peaking at
about 1 GHz in frequency (O'Dea 1998), and II) Compact Symmetric 
Objects (CSO) which are characterised by their small size ($<500$ pc) and 
two-sided radio structure, e.g. having jets and lobes on both sides of a 
central core (Wilkinson et al. 1994).  
Clearly, samples of GPS sources and CSOs are 
selected in very
different ways.  GPS sources are selected on their broadband radio
spectra, while CSOs are selected on their multi-frequency
milli-arcsecond morphology. Therefore studies of these objects have
mostly been presented separately.  However, a significant overlap
between the two classes of sources exists.  GPS sources optically
identified with galaxies are most likely to possess compact symmetric
radio morphologies (Stanghellini et al. 1997a, 1999), 
and the large majority of CSOs exhibit a gigahertz-peaked spectrum.  
The large but not complete overlap between
these two classes of source is most likely caused by the synchrotron
self-absorbed mini-lobes, located at the extremities of most CSOs,
being the main contributors to the overall radio spectrum, and
producing the peak at about 1 GHz in frequency.

\subsection{Evidence for the young nature of GPS sources and CSOs.}

Since the early discovery of GPS sources, it has been speculated
that these were young objects (Shklovsky 1965, Blake 1970). 
However, a commonly 
discussed alternative to them being young was that they are small 
due to confinement by a particularly dense and clumpy interstellar
medium that impedes the outward propagation of the jets 
(van Breugel, Miley \& Heckman 1984; O'Dea, Baum \& Stanghellini 1991).
This latter hypothesis now looks less likely since recent observations show 
that the surrounding media of peaked spectrum sources are not significantly
different from large scale radio sources, and insufficiently dense to 
confine these sources.
The most compelling evidence for youth is found in observations of the 
propagation velocities of the hot spots of several GPS/CSOs
(Owsianik \& Conway 1998; Owsianik, Conway \& Polatidis 1998;
Tschager et al. 1999). They all appear to have separation velocities
of typically $\sim 0.2h^{-1}c$, indicating a
dynamical age of $\sim 10^3$ year and clearly showing that these are
indeed young objects. Recent measurements of the
high frequency breaks in  Compact Steep Spectrum (CSS)
sources, indicate that these somewhat larger objects have radiative ages
in the range of $10^3$ to $10^5$ years (Murgia et al. 1999).

\subsection{Current views on Radio Source Evolution}

Observational constraints on the luminosity evolution of radio sources
mainly come from the source density in the power - linear size $(P-D)$ 
diagram (Shklovsky 1963). It was found that sources with large sizes
($D>1$ Mpc, eg. Schoenmakers 1999) and high radio luminosities ($P>10^{26}$ W/Hz at 178 MHz) 
are rare, suggesting that the luminosities of sources should decrease quickly
with linear sizes approaching 1 Mpc. Several authors have compared the 
number densities of GPS and CSS sources  with those of large radio 
sources to investigate the luminosity evolution of young radio sources
(Fanti et al. 1995; Readhead et al. 1996; O'Dea \& Baum 1997).
Fanti et al. (1995) argue that the luminosities of CSS sources decrease by
a factor of $\sim 10$ as they evolve to extended objects. Readhead et al. 
(1996) find a factor of 8 decrease in luminosity as a source expands from 
500 pc to 200 kpc in overall size. Taking into account their CSO statistics,
they find that the luminosity evolution in the phases CSO-MSO-LSO
(MSO = Medium Symmetric Object; LSO = Large Symmetric Object), i.e. 
from 10 pc to 150 kpc, is consistent with a single power-law luminosity
evolution. This conclusion is not supported by O'Dea \& Baum (1997), 
who found that GPS and CSS sources must decrease in luminosity at a 
faster rate than the classical 3CR doubles. 
Blundell, Rawlings \& Willott (1999) showed that any radio source evolution
involving a decrease in luminosity with time would, at the highest redshifts,
result in a bias towards young sources in flux density limited samples.
Since this effect is only important at $z>2-3$, it is not relevant
to the analysis presented in this paper.

Several GPS and CSOs (eg. 0108+388; Baum et al. 1990) exhibit low
level, steep spectrum, extended emission on arcsecond scales, which seem to
be relics of much older radio-activity. These objects are often classified
as being intermittent or re-occurent, and therefore not as young objects.
However, the components related to their gigahertz-peaked spectra and
CSO morphologies are certainly young. 
The presence of faint relic emission
only indicates that the active nucleus has been active before, and may
constrain the typical timescale and frequency of such events. 
Based on the current knowledge of the formation of massive 
black-holes in the centers of galaxies, 
it is unlikely that the central engine itself is young, but just the 
radio source (Richstone et al. 1998).

It is unclear whether all young sources actually evolve to large extended 
objects. Some, or even the majority, may be short-lived phenomena 
due to a lack of significant fuel (Readhead 1994). The possible existence
of these objects can have a large influence on the source statistics of young 
radio sources.

\section{Samples of GPS and CSS sources}

In this paper, we present a study of the evolution of young 
radio sources from the analysis of three samples of faint and bright
GPS and CSS radio sources:
The faint GPS sample from WENSS (Snellen et al. 1998a), the 
bright (Stanghellini et al. 1998) GPS sample, and the Fanti et al. (1990)
CSS sample. As discussed in the companion paper (Snellen, Schilizzi \& van
Langevelde, 2000), we do not regard GPS quasars to be related to their
galaxy counterparts; in the following
we choose not to use the quasars for
further analysis and only concentrate on the GPS and CSS galaxies.
Unless stated otherwise, we will assume that {\it all} GPS galaxies
evolve to large scale radio galaxies, and that {\it all} large scale radio 
galaxies were once GPS galaxies. 

\subsection{The faint GPS sample from WENSS}

The selection of this sample has been described in detail 
in Snellen et al. (1998a). Candidate GPS sources were selected in two ways;
i) with inverted spectra between 325 MHz and 625 MHz in WENSS
(Rengelink et al. 1997), and ii) with inverted 
spectra between WENSS 625 MHz and Greenbank 5 GHz (Gregory \& Condon 1991). 
The sources are located in two regions of the survey; one at $15^h < \alpha < 
20^h$
and $58^\circ< \delta < 75^\circ$, which is called the {\it mini-survey} region
(Rengelink et al. 1997), and the other at $4^h00^m < \alpha < 8^h30^m$ and
$58^\circ< \delta < 75^\circ$. Additional observations at 1.4, 5, 8.4 and 15
GHz were carried out with the WSRT and the VLA, yielding a sample of 47
genuine GPS sources with peak frequencies ranging from 500 MHz to more than 15
GHz, and peak flux densities ranging from $\sim30$ to $\sim900$ mJy.
This sample has been imaged in the optical and near-infrared, resulting in
an identification fraction of $\sim$ 87 \% (Snellen et al. 1998b, 1999).
All the galaxies in the sample were used for the morphological evolution 
study. The redshifts of the majority of the objects 
had to be estimated from their optical magnitudes, using
the well determined Hubble diagram for GPS galaxies 
(Snellen et al. 1996, O'Dea et al.1996). 
Some, assumed to be galaxies, have only faint lower limits to their 
magnitudes. For these a redshift of z=1.5 was assumed.

The overall angular sizes were 
measured from the VLBI observations as the maximum angular 
separation of components or the angular size for single component source
(see companion paper)
Their 5 GHz radio power was determined, assuming $H_o=50$ km sec$^{-1}$ Mpc$^{-1}$
and $\Omega_o=1$. For a few sources, the rest-frame peak frequency was above 5 GHz.
The radio power of these galaxies was corrected for the spectral turnover
by extrapolating their optically thin spectrum to rest-frame 5 GHz.
In the most extreme case (B0752+6355), this correction is $<20\%$.
B0531+6121 was omitted from the sample since it does not
have a genuine GPS spectrum.

For the luminosity evolution study to be discussed in section 4, 
it is crucial to have a good understanding of the selection effects. 
We therefore applied 
more strict constraints than in the original sample. Only the GPS sources 
which have inverted spectra between 325 MHz and 5 GHz, and with 
flux densities of $>20$ mJy at 325 MHz, 14 in total, were selected.

All the 26 galaxies in the sample are given in table \ref{WENSS}.
Column 1 gives the B1950 name, column 2 indicates whether the 
source is in the complete sub-sample or not, column 3 gives the 
(estimated) redshift, column 4 gives the observed peak frequency, 
column 5 the observed peak flux density,
column 6 the rest-frame 5 GHz radio power, and column 7 the overall
angular size.   

\begin{table}
\setlength{\tabcolsep}{1mm}
\centerline{
\begin{tabular}{ccrcrcr}\hline
Name&C&z&$\nu_{peak}$&$S_{peak}$&$P_{5GHz}$&$\theta$\\
    & & &  (GHz)     &  (mJy)   &  (Log$\frac{W}{Hz}$)  &  (mas)\\ \hline
B0400+6042&+&1.5$^2$&1.0& 184& 26.8 & 4.4\\
B0436+6152&+&1.5$^2$&1.0& 237& 26.9 &17.1\\
B0535+6743&+&1.5$^1$&5.7& 192& 27.2 & 4.0\\
B0539+6200&+&1.4$^1$&1.9& 129& 26.6 & 6.1\\
B0552+6017&-&1.5$^2$&1.0&  50& 26.2 &12.6\\
B0557+5717&-&1.2$^1$&1.1&  69& 26.2 & 6.5\\
B0752+6355&-&0.9$^1$&6.4& 314& 26.7 & 4.6\\
B0759+6557&-&1.5$^1$&1.7&  46& 26.2 & 8.0\\
B0830+5813&+&0.093  &1.6&  65& 24.1 &$<4.3$\\
B1525+6801&+&1.1$^1$&1.8& 163& 26.5 &22.4\\
B1551+6822&+&1.3$^1$&1.5&  52& 26.1 & 2.5\\
B1557+6220&+&0.9$^1$&2.3&  49& 25.8 &$<4.4$\\
B1600+7131&+&1.5$^2$&1.7& 346& 27.0 &22.7\\
B1620+6406&-&1.2$^1$&2.2&  47& 25.9 &14.2\\
B1622+6630&+&0.201  &4.0& 363& 25.5 &$<2.9$\\
B1639+6711&-&1.5$^2$&1.0&  68& 26.4 &$<4.1$\\
B1655+6446&+&1.5$^2$&1.0&  69& 26.3 &24.8\\
B1657+5826&-&1.1$^1$&0.5&  64& 26.0 &27.8\\
B1807+5959&-&1.0$^1$&1.0&  47& 25.9 &13.0\\
B1807+6742&-&1.5$^2$&0.8&  54& 26.2 & 6.9\\
B1808+6813&-&1.1$^1$&1.3&  42& 25.9 & 3.5\\
B1819+6707&-&0.220  &0.8& 338& 25.5 &19.2\\
B1841+6715&+&0.486  &2.1& 178& 26.0 & 6.1\\
B1843+6305&-&1.5$^2$&1.9&  75& 26.4 & 9.5\\
B1942+7214&+&1.1$^1$&1.4& 233& 26.7 &31.5\\
B1946+7048&+&0.101  &1.8& 929& 25.4 &31.9\\ \hline
\end{tabular}}
\centerline{
\begin{tabular}{l}
$^1$ estimate from optical magnitude\\
$^2$ assumed at z=1.5\\
\end{tabular}}
\caption{\label{WENSS} The GPS galaxies from the faint WENSS sample.
The second column (+/$-$) indicates whether a source is part of the 
complete sub-sample or not.}
\end{table}

\subsection{The bright Stanghellini et al. GPS sample}

A sample of radio bright GPS sources has been constructed by
Stanghellini et al. (1998) from GPS candidates selected from the K\"uhr 
et al. (1981)
 1 Jy catalogue, with declination 
$>-25^{\circ}$ and galactic latitude $|b|>10^\circ$.
Stanghellini et al. supplemented this data set with multi-frequency 
observations from the VLA, WSRT and data from the literature, and
selected sources with a turnover frequency between 0.4 and 6 GHz, and 
an optical thin spectral index $\alpha_{thin}<-0.5$ at high frequency.
The final complete sample consists of 33 GPS sources, of which 19 are
optically identified with galaxies. 
Four galaxies do not have a spectroscopic redshift. Their redshifts were
estimated from their optical magnitudes, in the same way as for 
galaxies in the WENSS-sample. Their rest-frame radio power at 5 GHz has
also been calculated in the same way as for the objects in the WENSS sample.

All the galaxies in the sample are given in table \ref{BRIGHT}.
Column 1 gives the B1950 name, column 2 the 
redshift, column 3 the observed peak frequency, 
column 4 the observed peak flux density,
column 5 the rest-frame 5 GHz radio power, and column 6 the overall
angular size. Column 7 gives the reference for the angular size.

\begin{table}
\caption{\label{BRIGHT}The complete sample of Bright GPS galaxies
from Stanghellini et al. (1998)}
\setlength{\tabcolsep}{1mm}
\centerline{
\begin{tabular}{crcrcrr}\hline
Name&z&$\nu_{peak}$&$S_{peak}$&$P_{5GHz}$&$\theta$&Refs.\\
    & &  (GHz)     &  (Jy)   &  (Log$\frac{W}{Hz}$)  &  (mas)\\ \hline
0019$-$000&0.305&0.8&3.5&26.7& 70&5\\
0108+388  &0.669&3.9&1.3&27.3&  6&6\\
0316+161  &1.2  &0.8&9.6&28.2&300&3\\
0428+205  &0.219&1.0&4.0&26.6&250&3\\
0500+019  &0.583&2.0&2.5&27.3& 15&1\\
0710+439  &0.518&1.9&2.1&27.1& 25&7\\
0941$-$080&0.228&0.5&3.4&26.4& 48&8\\
1031+567  &0.459&1.3&1.9&26.9& 33&9\\
1117+146  &0.362&0.5&3.9&26.8& 90&10\\
1323+321  &0.369&0.5&7.0&27.1& 60&3\\
1345+125  &0.122&0.6&8.9&26.3& 85&1\\
1358+624  &0.431&0.5&6.6&27.1& 80&3\\
1404+286  &0.077&4.9&2.8&25.8&  8&4\\
1600+335  &1.1  &2.6&3.1&27.9& 60&3\\
1607+268  &0.473&1.0&5.4&27.2& 49&8\\
2008+068  &0.7  &1.3&2.6&27.3& 30&2\\
2128+048  &0.990&0.8&4.9&27.8& 35&1\\
2210+016  &1.0  &0.4&4.5&27.7& 80&1\\
2352+495  &0.237&0.7&2.9&26.5& 70&7\\ \hline
\end{tabular}}
\centerline{
\begin{tabular}{ll}
Refs for angular sizes:\\
1) Stanghellini et al. (1997a)& 2) Stanghellini et al. (1999)\\
3) Dallacasa et al. (1995)& 4) Stanghellini et al. (1997b)\\
5) Hodges, Mutel \& Phillips (1984)& 6) Owsianik et al. (1998)\\
7) Wilkinson et al. (1994)& 8) Dallacasa et al. (1998)\\
9) Taylor, Readhead \& Pearson (1996)& 10) Bondi et al. (1998).\\
\end{tabular}}
\end{table}

\subsection{The Fanti et al. CSS sample}

The sample of CSS sources used in this paper is 
from Fanti et al. (1990). They constructed a sample without spectral bias 
by integrating the 3CR sample with sources from the Peacock and Wall
sample (1982) which would be stronger than 10 Jy at 178 MHz, if
corrected for low frequency absorption by extrapolation of the straight 
high-frequency part of the spectrum. All sources were included
with projected linear size $<15$ kpc, (corrected) flux density at
178 MHz $>$10 Jy, and with Log $P_{178}>$ 26.5, in a well defined
area of sky ($|b|>10^\circ$, $\delta>10^\circ$). A number of sources, 
which are included in the Stanghellini et al. sample are
omitted from the Fanti et al. sample to avoid duplication.
The remaining CSS galaxies are listed in table \ref{CSS}.
 
\begin{table}
\caption{\label{CSS} The complete sample of CSS galaxies 
from Fanti et al. (1990).}
\setlength{\tabcolsep}{1mm}
\centerline{
\begin{tabular}{lrcrcrr}\hline
Name&z&$\nu_{peak}$&$S_{peak}$&$P_{5GHz}$&$\theta$&Refs.\\
        & &  (GHz)     &  (Jy)   &  (Log$\frac{W}{Hz}$)  &  ($''$)\\ \hline
3C49    &0.62&0.12   &  11  &27.2 & 1.0 &1\\
3C67    &0.31&0.05   &  10  &26.6 & 2.5 &1\\
3C93.1  &0.24&0.06   &  10  &26.3 & 0.6 &2\\
0404+76 &0.59&0.60   &   6  &27.6 & 0.1 &2\\
3C237   &0.88&0.05   &  40  &27.9 & 1.2 &1\\
3C241   &1.62&0.04   &  17  &27.8 & 0.8 &1\\
3C268.3 &0.37&0.08   &  11  &26.9 & 1.3 &1\\
3C299   &0.37&0.08   &  13  &26.8 &11.5 &1\\
3C303.1 &0.27&0.10   &  10  &26.4 & 2.0 &1\\
3C305.1 &1.13&0.09   &  10  &27.5 & 2.8 &1\\
3C318   &0.75&$<0.04$&  20  &27.3 & 0.8 &1\\
3C343.1 &0.75&0.25   &  13  &27.6 & 0.3 &1\\
3C346   &0.16&$<0.04$&  10  &26.2 & 2.3 &1\\
1819+39 &0.80&0.10   &   7  &27.6 & 0.5 &2\\
3C454.1 &1.84&$<0.04$&  10 &27.8 & 1.6 &1\\
\end{tabular}}
\centerline{
\begin{tabular}{l}
Refs for angular sizes:\\
1) Spencer et al. 1989\\
2) Dallacasa et al 1995\\
\end{tabular}}
\end{table}

\section{The spectral turnovers and the morphological evolution of young 
radio sources}

\begin{figure*}
\psfig{figure=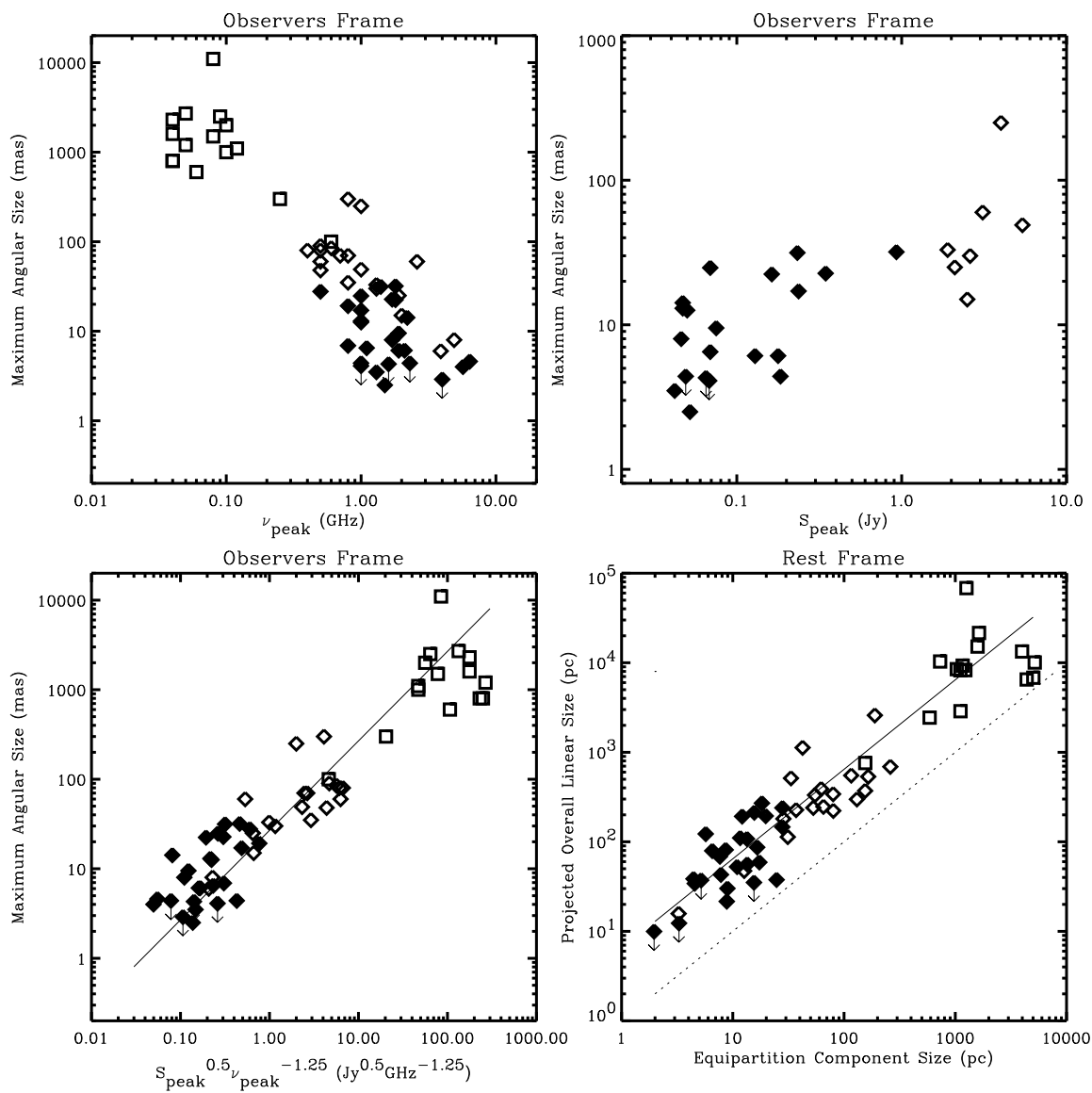,width=16cm}
\caption{\label{morph} Correlations found between the spectral turnover and 
the overal size in samples of GPS and CSS sources. Squares, open
and filled diamonds indicate objects from the CSS, bright and faint GPS
sample respectively.
(top left) The maximum angular size versus peak frequency.
(top right) The maximum angular size versus peak flux density for
sources with $0.8<\nu_p<3$.
(bottom left) The maximum angular size versus  $S_p^{1/2} \nu_p^{-5/4}$.
The line indicates a linear correlation between the two parameters.
(bottom right) The overall linear size versus the equipartition component
size. The distance of the data-points to the dotted line indicates
the component to overall size ratio.}
\end{figure*}

Early measurements of the angular sizes of GPS and CSS sources using VLBI
strongly suggested that their spectral turnovers are caused by
synchrotron self absorption (SSA, Jones, O'Dell \& Stein 1974; Hodges, Mutel
\& Phillips 1984; Mutel, Hodges \& Phillips 1985)
It was realised by Jones, O'Dell \& Stein (1974) that if the optical depth
due to SSA was less than unity at the spectral peak-frequency of these 
sources,  
lower magnetic fields, far from equipartition, would be present 
which should result in detectable self-compton radiation.
Fanti et al. (1990) showed that there is a strong anti-correlation between 
the linear size and the turnover frequencies of CSS sources, as expected
for SSA. However, more recently it was suggested by Bicknell, Dopita \&
O'Dea (1997) that such a correlation can also be explained by a particular
model in which these sources undergo free-free absorption by ionised gas
surrounding the lobes. 
In addition, Kuncic, Bicknell \& Dopita (1998) argued that in addition to
free-free absorption, induced Compton scattering will also have an important 
effect in forming the spectral peak. As a result, 
it opened up the debate again 
about the cause of the spectral turnovers in CSS and GPS sources.

The combination of bright and faint GPS and CSS samples as presented here,
gives us a unique opportunity to carefully investigate the correlation between
size and spectral peak. 
Not surprisingly, we confirm the anti-correlation between peak frequency 
$\nu_p$ and maximum angular size $\theta_{max}$ (see figure \ref{morph}, top
left panel).
However, in addition we find a correlation between peak flux density $S_p$ and 
$\theta_{max}$. This is shown in the top right panel of figure \ref{morph}.
Note that only sources from the bright and faint GPS samples are
with $0.8<\nu_p<3$ GHz are plotted here. This is necessary, since the 
peak flux densities are correlated with the
 peak frequencies, which would erroneously result in a correlation between
peak flux density and angular size.

From SSA theory, it is expected that the angular size $\theta$ of a radio 
source is proportional to (Kellerman \& Pauliny-Toth,1981):
\begin{equation}\label{eq1}
    \theta \propto B^{1/4} S_p^{1/2} (1+z)^{1/4} \nu_p^{-5/4}
\end{equation}
where $B$ is the magnetic field strength and $z$ the redshift. 
Note that $\theta$ is
only weakly dependent on both $B$ and $z$. Most remarkably, the strength
and signs of the correlations between $v_p$, $S_p$ and $\theta_{max}$ as 
shown in figure \ref{morph} are exactly as expected from equation \ref{eq1}.

The overall angular size, $\theta_{max}$
(eg. the distance between the two mini-lobes), is used in the analysis above, 
but $\theta$ in equation \ref{eq1} corresponds to the size of the radio 
{\it components} which are dominant at the peak-frequency (the mini-lobes).
Therefore, these correlations have implications for the 
morphological evolution of these radio sources. 
The lower left panel of figure \ref{morph} shows the maximum angular size
as function of $S_p^{1/2}\nu_p^{-5/4}$. The solid line indicates the 
best linear fit. The dependence of this relation on redshift is 
proportional to  $(1+z)^{1/4}$, which in any case is smaller than 
$<20\%$ and negligible for our z-range. 
Therefore the same relation is expected in the rest-frame of 
the objects. In the rest-frame, we can solve for the magnetic field 
$B$ by assuming equipartition. For this we use the equation derived by 
Scott \& Readhead (1977) assuming an optically thin spectral index 
$\alpha=-1$,
\begin{equation}
L = 3.5\times (1-(1+z)^{-1/2})^{-1/17}(1+z)^{1/2}S_p^{8/17}\nu_p^{33/34}
\end{equation} 
where $L$ is the equipartition component size. The projected linear size
is shown as function of the equipartition component size in the lower right 
panel of figure \ref{morph}. The dashed line indicates the dependence for which
both quantities are the same. The solid line is the best linear least-squares 
fit, indicating a ratio of overall size to component size of $5-6$, throughout
the samples of faint and bright GPS and CSS galaxies.
This means that if GPS sources evolve into CSS sources, their ratio of 
component size to overall linear size remains constant, implying a 
self-similar evolution.

Note that the main difference between the lower left and right panels
of figure \ref{morph} is that in the first a constant magnetic field is 
assumed, and in the second an equipartition magnetic field. It appears
that the first correlation is slightly flatter than expected for self-similar
evolution. Indeed the ratio of the component to overall angular size 
is on average a factor 2 smaller for the CSS sample than for the GPS samples,
while these ratios are virtually the same assuming an equipartition magnetic 
field. This may indicate that young radio sources stay in equipartition 
while evolving in a self-similar way. This would require that the 
magnetic fields in CSS sources are typically a factor $\sim20$ lower than
in GPS sources. The linear correlation itself only indicates a constant
ratio between the magnetic field and particle energies.
This constant does not have to be equal to 
unity, as required for equipartition.
However, for a ratio of unity in energies, the bottom right panel of 
Figure \ref{morph} requires a ratio of overall to component size 
of typically $5-6$, which is close to the result seen in VLBI 
observations. This means that the energy ratio is not only constant, 
but also close to unity, which indicates that equipartition probably holds.

Figure \ref{morph} demonstrates that the data is consistent with
a combination of SSA, equipartition, and self-similar growth. 
It is not obvious that the same correlation
should apply for free-free absorption. Although other more complicated 
combinations of mechanisms such as free-free absorption with
induced Compton scattering (Kuncic, Bicknell \& Dopita, 1998)
may also fit the data, the simplest explanation by far is to assume 
that SSA, equipartition and self-similar source growth all individually 
hold. We therefore believe  that SSA is indeed the cause of the spectral 
turnovers in GPS and CSS sources.

It may not be surprising that young radio sources evolve in a self-similar 
way. Leahy and Williams (1984) showed that the cocoons of FRII sources of
very different physical size had similar axial ratios. More recently, 
Subrahmanyan, Saripalli \& Hunstead (1996) found very similar ratios for 
sources of linear sizes above 900 Kpc, also suggesting that radio sources 
evolve in a self-similar way. An analytical model for radio sources with 
pressure confined jets developed by Kaiser \& Alexander (1997) 
shows that the properties of the bow shock and of the surrounding gas 
{\it force} the sources to grow in a self-similar way, provided that the 
density of the surrounding gas falls off less steeply than $1/r^2$.

\section{The luminosity evolution of young radio sources}

\begin{figure}
\psfig{figure=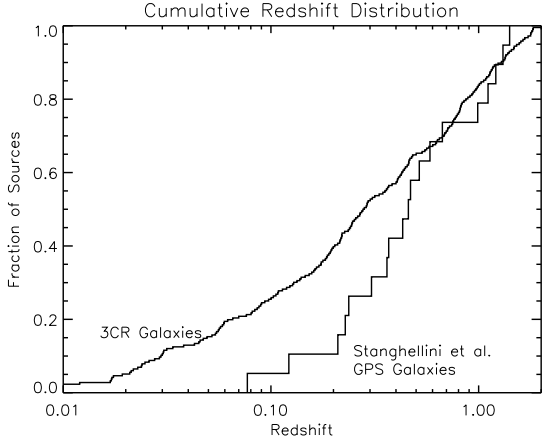,width=8cm}
\caption{ \label{reddis} The cumulative redshift distribution for 3C galaxies
and GPS galaxies from the Stanghellini et al. sample.}
\end{figure}

The number count statistics and linear size distributions  
used in studies to constrain the luminosity evolution of radio sources, have
all been averaged over a wide redshift range and only include the brightest 
objects in the sky (Fanti et al. 1995, Readhead et al. 1996, O'Dea \& Baum 1997). However, in flux density limited
samples, the redshift distribution of GPS galaxies is significantly different
from that of large size radio galaxies (see figure \ref{reddis}). 
This suggests that the interpretation of the number count statistics is not 
straightforward.
Note that given the expected luminosity evolution as sources evolve in size,
many of the present day GPS sources will have FRI luminosities. 
It is therefore assumed that GPS galaxies evolve into both FRI and FRII 
sources.

The bias of GPS galaxies towards higher redshifts than large size radio 
galaxies itself provides an important clue about the luminosity evolution of 
radio sources. 
It implies that GPS galaxies are more likely to have higher
radio power than extended objects in flux density limited
samples. If GPS and large size radio sources are identical objects, 
observed at different ages, their cosmological density evolution, for 
example 
their birth rate as function of redshift, should be the same. Since their 
lifetimes are short compared to the Hubble time, the redshift distributions 
of the GPS galaxies, and the objects they evolve to, should 
also be the same. The bias of GPS sources towards higher redshifts and 
radio powers 
therefore implies that their luminosity function must be flatter than
that of large size radio sources. We argue that the luminosity evolution
of the individual objects strongly influences their collective luminosity 
function, and propose an evolution scenario in which GPS sources 
increase in luminosity and large size sources decrease in luminosity with time
(see section 4.1).
In the simplified case, in which source to source variations in
the surrounding medium can be ignored,
the luminosity of a radio source depends only on its age and jet power.
Consider first the luminosity function
 of large size sources. It is expected that 
large size sources decrease in luminosity with age (see section 4.1). 
Therefore high luminosity sources will tend to be biased towards objects 
with both small ages and high jet powers. The intrinsic space density for 
high power jet sources will of course tend to be small.
Furthermore, for a given jet power there are fewer young sources than
old sources, simply because sources spend only a small fraction of their time 
being young. The result is a very low space density of large size sources
of high power. In contrast, large size sources of low power are biased to 
be both old and with low jet power, both common conditions, hence the space 
density of large size sources with low power is much higher than that for 
those with high power, and the luminosity function for large size sources 
is steep. In contrast, the luminosity of GPS sources is expected to 
increase with sources age (see section 4.1). High luminosity GPS sources
are therefore biased to be old and of high jet power, while low luminosity
objects are biased to be young and of low jet power. Instead of 
reinforcing each other as in the case for large size sources, for GPS sources
the age and jet power space density biases partly counteract.
The result is a much less difference in the space density of 
low and high power GPS sources and hence a much flatter luminosity 
function for GPS sources.

In the next section we will show that 
the luminosity evolution as proposed is expected for a ram-pressure confined,
self similarly evolving
radio source in a surrounding medium with a King-profile density. 
In the inner parts of the King profile, the density of the medium is 
constant and the radio source builds up its luminosity (eg. Baldwin 1982), 
but after it grows
large enough the density of the medium declines and the luminosity of the radio
source decreases. 

In section \ref{liflaf} we will show how the luminosity evolution 
of the individual sources  modifies the luminosity function, and in section
\ref{loclumfun}, the local luminosity function of GPS sources is constructed
and compared with that of large size radio sources.

\subsection{A self-similar evolution model}

An important parameter in evolution models of radio sources is the 
density profile of the surrounding medium. In general, X-ray 
observations of nearby ellipticals have shown that their 
ISM are well fitted by a King profile distribution (Trinchieri et al. 1986):
\begin{equation}
\rho(r) = \rho_0 \biggl[ 1+\biggl(\frac{r}{r_c}\biggr)^2\biggr]^{-\beta/2}
\end{equation}
where $\rho$ is the density of the medium as function of distance to 
the centre of the host galaxy $r$, $r_c$ is the core radius, 
and $\beta$ the slope parameter.
Typical core radii in giant ellipticals
are observed to be $r_c=500-1000$ pc (Trinchieri et al. 1986), 
For simplicity, we treat the two regimes separately: 1) The GPS phase 
at $r<r_c$ where the density of the medium, $\rho_{ism}$, 
is assumed to be constant. 2) The large size (LS) phase at $r>r_c$ where
$\rho_{ism} \propto r^{-\beta}$.

If the thrust of the radio jet is balanced by the ram-pressure
of the surrounding medium, the growth of the radio source is equal to
\begin{equation}\label{Meq1}
dr/dt \propto \biggl(\frac{P_J}{\rho_{ism}(r) A}\biggr)^{1/2}
\end{equation}
where $dr/dt$ is the propagation 
velocity of the hot-spots, $P_J$ is the jet power, 
and $A$ is the cross-sectional area (Begelman 1996).
In the previous section we showed that young radio sources
seem to evolve in a self-similar way. Since it is in close 
agreement with the theoretical work of Kaiser \& Alexander (1997), we will 
assume self-similar evolution, with $A \propto r^2$.
Note however, that in the work by Kaiser \& Alexander (1997), the 
cross-sectional area of the jet grows slightly more slowly with size,
which is therefore not completely self-similar, but 
allowing the expansion of the bow-shock and cocoon to be fully self-similar
(Kaiser 2000, private communications). 
Here, by assuming $A\propto r^2$, this will not be the case, 
but differences are small and for simplicity's sake we use this anyway.

From integrating eq. \ref{Meq1} it follows that in the GPS phase a source grows
in linear size with time as $t^{1/2}$, assuming that the jet-power is 
constant with time. The average internal density of the radio source, 
$\rho_i$, is proportional to $P_Jt/V$, where $V$ is the volume
of the radio source which is proportional to $r^3$. 
Hence,
$\rho_i\propto r^{-1}$, indicating that the radio emitting 
plasma expands proportionally to its linear size, $r$, and that 
expansion losses have to be taken into account. If the energy
spectrum of the electrons is $n(E)=n_oE^{-\gamma}$, $n_o$ varies 
proportionally to $r^{-4/3}$ for $\gamma=2$ ($\alpha=-0.5$, Moffet, 1977).
We will assume that the radio source is in equipartition, so
that $n_o \propto B^2$, where $B$ is the magnetic field.
The radio power $L_\nu$
at a particular frequency in the optically thin part of the spectrum
scales as
\begin{equation}
L_\nu \propto n_o^{7/4} V \propto P_J^{7/8}r^{2/3} 
\end{equation}
for $\gamma=2$. Hence radio sources increase in luminosity in the 
GPS phase.

\begin{figure}
\psfig{figure=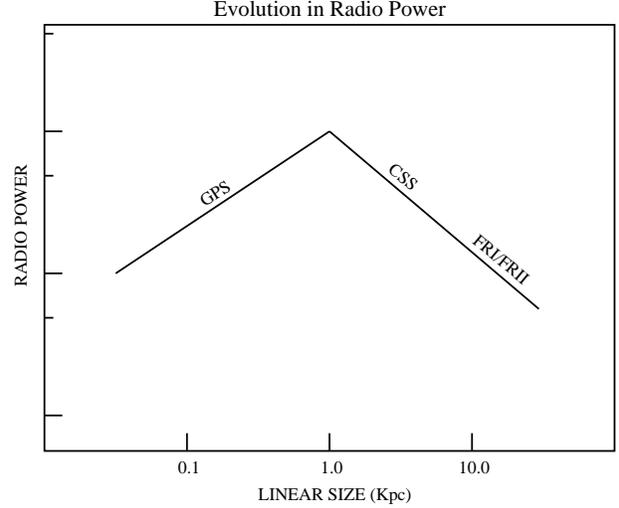,width=8cm,angle=-90}
\caption{\label{lumevol} The evolution in radio power as function of 
linear size for a self-similar evolving, ram-pressure confined radio source
in a surrounding medium with a King-profile density. }
\end{figure}

In the LS phase, radio sources grow as $t^{2/(4-\beta)}$, and the
density of the radio emitting plasma varies as $\rho_i\propto r^{-\frac{\beta+2}{2}}$,
Taking into account expansion losses in a similar way
as for the GPS phase, this means that
$n_o \propto r^{-4\frac{\beta+2}{6}}$, and under equipartition conditions,
\begin{equation}
L_\nu \propto P_J^{7/8}r^{\frac{2}{3}-\frac{7}{6}\beta}
\end{equation}
Hence radio sources in the LS phase decrease in radio luminosity.
Note that we do not take synchrotron losses and losses due to scattering 
of the CMB into account, 
which may influence the LS phase and cause sources to decrease faster
in luminosity with time.
The schematic evolution in radio power of a radio source according to this 
model is shown in figure \ref{lumevol}. 

\begin{figure}
\psfig{figure=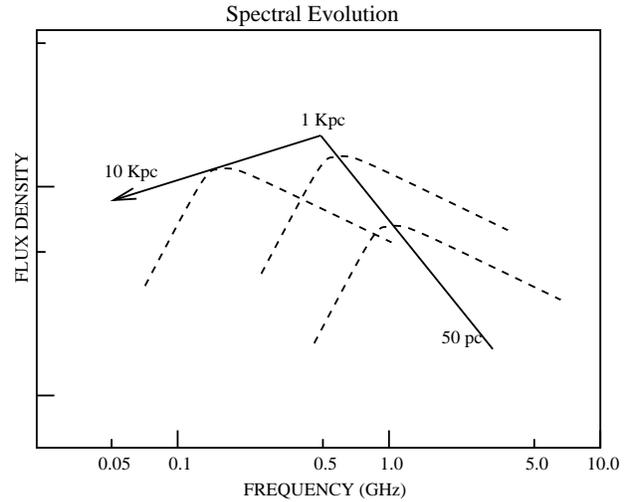,width=8cm,angle=-90}
\caption{\label{evoltracks} An evolutionary track  
for a self-similar evolving, ram-pressure confined radio source
in a surrounding medium with a King-profile density}
\end{figure}

It is interesting to determine what the expected evolution in peak frequency
and peak flux density is for a radio source in the GPS phase and the LS phase.
A source will become optically thick at a frequency where $\kappa_\nu l \approx
1$, where $\kappa_\nu$ is the absorption coefficient and $l$ the pathlength 
through the radio plasma.  For synchrotron self absorption, assuming 
equipartition and self-similar evolution, this means that,
\begin{equation}
\kappa_\nu l \ \propto \ n_o B^{2} \nu_p^{-3} r \ \propto \ n_o^2 \nu_p^{-3} 
r=1
\end{equation}
for $\gamma = 2$ (Moffet, 1977). 
The optically thin radio power, as determined above,
will be frequency dependent and proportional to  
$P_\nu \propto S_p\nu_p^{1/2}$. Therefore in the GPS phase,
\begin{equation}
\nu_p \propto r^{-5/9}, \ \ S_p \propto r^{17/18}, \ \
S_p \propto \nu_p^{-17/10}
\end{equation}
In the LS phase,
assuming $\beta=1.5$, which is a typical value based on 
observations of X-ray halos (Trinchieri et al. 1986),
\begin{equation}
\nu_p \propto r^{-11/9}, \ \ S_p \propto  \nu_p^{17/44}
\end{equation}
From figure \ref{morph} we can estimate the transition between the two
phases (500-1000 pc) to occur at $\nu_p \approx 100-500$ MHz  
The evolutionary tracks are shown in figure \ref{evoltracks}.

\subsection{Luminosity evolution and the luminosity function.\label{liflaf}}

In this section the influence of the luminosity evolution of 
the individual objects on the slope of their collective luminosity 
function is derived. 
We will ignore source to source variations in the surrounding
medium and use the radio-size dependent luminosity evolution 
as derived in the previous section.
\begin{figure}
\psfig{figure=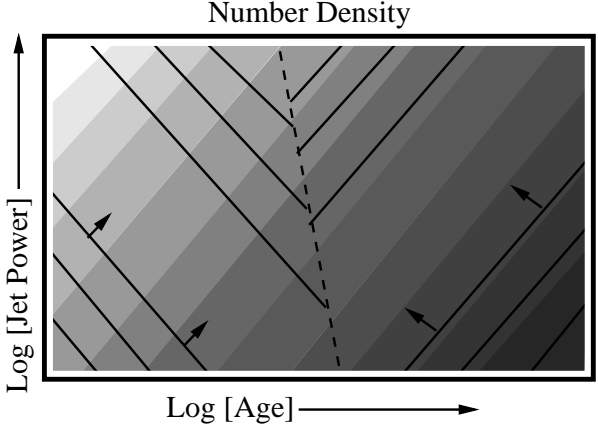,width=8cm,angle=-90}
\caption{\label{integral} Schematic representation of the source density
as function of age and jet power. The lines indicate sources with identical
luminosities. The greyscales indicate the source density for a particular jet 
power and age. The dashed line indicates the border between the GPS phase and 
the LS phase.}
\end{figure}

Suppose that the comoving number density of sources with a jet 
power $N(P_J)$ is a power-law distribution
\begin{equation}
N(P_J) \propto P_J^\delta
\end{equation}
between $P_-$ and $P_+$, 
and the sources have a flat distribution of ages below a certain maximum 
age, then the source density as function of age and jet power is 
represented by the grey scales in figure \ref{integral}.
The radio power of a source, $L_\nu$, can be parameterised as 
\begin{equation}
L_\nu \propto P_J^\kappa r^\epsilon
\end{equation}
where $\epsilon=2/3$ and $\epsilon=-13/12$ in the GPS phase and LS phase
respectively, and $\kappa=7/8$, as derived in the previous section.
A line-integral over a solid line in figure \ref{integral} gives
the total number of sources in the volume with a particular 
luminosity. It can be seen that in the GPS phase, these lines 
are approximately perpendicular to the 
density gradient, indicating that a change in luminosity results
in only a small change in the number of sources. 
In the LS phase, they 
are parallel to the density gradient, and 
a change in luminosity results in a large change in the number of objects.
The luminosity function $ N(L)$ can be derived 
from,
\begin{equation}
N(L_\nu) = \frac{\delta}{\delta L_{*}} \iint\limits_{L_\nu(p,r)<L_{*}}
 N(p,r) 
dp dr
\end{equation}
where $p$ is the jet power and $r$ is the size of the radio source.
As can be seen in figure \ref{integral}, the integration limits of this 
equation are a different function of age and jet power, depending 
on the luminosity . The equation should be
solved separately for a high and low luminosity regime, in both the 
LS and the GPS phase. Since the border between the GPS and LS phase
is at a constant source size, $r_{*}$, it is better to integrate over
the source size $r$ than over the source age $t$.
 For the low luminosity regime in the GPS phase, 

\begin{equation}
N(L_\nu<L_*)=\int\limits_{P_{-}}^{P_{+}}\int\limits_0^{\frac{L_*^{1/\epsilon}}{p^{\kappa/\epsilon}}} 
p^{\delta-\frac{1}{2}}r dr dp = L_*^{\frac{2}{\epsilon}}\int\limits_{P_{-}}^{P_{+}}
p^{\delta-\frac{1}{2}-\frac{2\kappa}{\epsilon}}
dp
\end{equation}

where $N(L_\nu<L_*)$ are the total number of sources below a 
particular luminosity $L_*$, and therefore integrating over jet power $p$
and age $t$ gives,
\begin{equation}
N(L_\nu) \propto L_\nu^{2/\epsilon-1}
\end{equation}
For the high radio power regime in the GPS phase,
\begin{equation}\label{16}
N(L_\nu\!\!  > \!\!  L_*)=\!\!\int\limits_{\left(\frac{L_*}{P_+^\kappa}\right)^{\frac{1}{\epsilon}}}^{r_*}r \int\limits_{\left(\frac{L_*}{r^\epsilon}\right)^{\frac{1}{\kappa}}}^{P_{+}} 
p^{\delta-\frac{1}{2}} dp dr  \propto L_*^{(\delta+\frac{1}{2})\frac{1}{\kappa}}
\end{equation}
with a sharp cut-off near $L_\nu = P_+^\kappa r_*^{\epsilon}$. 
In this regime of radio power,
\begin{equation}
N(L_\nu)  \propto L_\nu^{(\delta+\frac{1}{2})\frac{1}{\kappa}-1}
\end{equation}
For the low luminosity regime in the LS phase, 
\begin{equation}
N(L_\nu<L_*)=\int\limits_{p_{-}}^{\left(\frac{L_*}{r_+^\epsilon}\right)^\frac{1}{\kappa}} p^{\delta-\frac{1}{2}}
\int\limits_{\left( \frac{L_*}{p^\kappa} \right)^{\frac{1}{\epsilon}}}
^{r_+} r^\frac{1}{2}  dr dp \propto L_*^{(\delta+\frac{1}{2})\frac{1}{\kappa}}
\end{equation}
with a sharp cut-off near $L_\nu=P_-^\kappa r_+^\epsilon$. In this regime of radio power,
\begin{equation}\label{18}
N(L_\nu) \propto L_\nu^{(\delta+\frac{1}{2})\frac{1}{\kappa}-1}
\end{equation}
For the high radio power regime in the LS phase,
\begin{equation}
N(L_\nu>L_*)=\int\limits^{\left(\frac{L_*}{P_+^\kappa}\right)^{\frac{1}{\epsilon}}}_{r_*} r^\frac{1}{2}
\int\limits^{P_+}_{\left(\frac{L_*}{r^\epsilon}\right)^\frac{1}{\kappa}}
 p^{\delta-\frac{1}{2}} dp dr \propto L_*^{3/2\epsilon}
\end{equation}
with a cut-off near  $L_\nu=P_+^\kappa r_*^\epsilon$. In this radio power regime,
\begin{equation}
N(L_\nu) \propto L_\nu^{3/2\epsilon-1}
\end{equation}

As can be seen from equations \ref{16} and \ref{18}, the slope of 
the luminosity function is expected to be the same in the 
high luminosity and low luminosity regimes for the GPS and 
LS phases respectively, since $\delta$ and $\kappa$ are independent of the 
age of the radio source. The low luminosity regime and the high
luminosity regime of the GPS phase and the LS phase are expected
to have a slope of $+2$ and $-2.4$ respectively for the proposed evolution
model.

\subsection{The Local Luminosity Function of GPS sources.\label{loclumfun}}

As is shown in the previous section, the comparison of the local luminosity 
function (LLF) of young and 
old radio sources can put strong constraints on the rise and 
decay of their radio luminosity. 
One would like to compare the LLF of GPS sources with the model derived in 
section 4.1 \& 4.2 directly. This is not possible due to the lack
of local GPS sources in present samples (the low local number density of 
GPS sources catalysed this discussion in the first place).
 For example, only 2 GPS 
galaxies in the Stanghellini et al. sample are at $z<0.2$.
However, since we assume that GPS sources evolve into 
large size sources and their lifetimes are short compared to 
cosmological timescales, their birth rate as function of redshift
 should be the same. Therefore
the cosmological evolution as determined for large scale radio sources can
be used to describe the cosmological evolution for GPS sources.
In this way, the GPS LLF can be estimated using the GPS galaxies at
all redshifts, which will be attempted in this section.
This estimated GPS LLF will then be compared with what is expected
from the model, as derived in section 4.2.

\begin{figure}
\psfig{figure=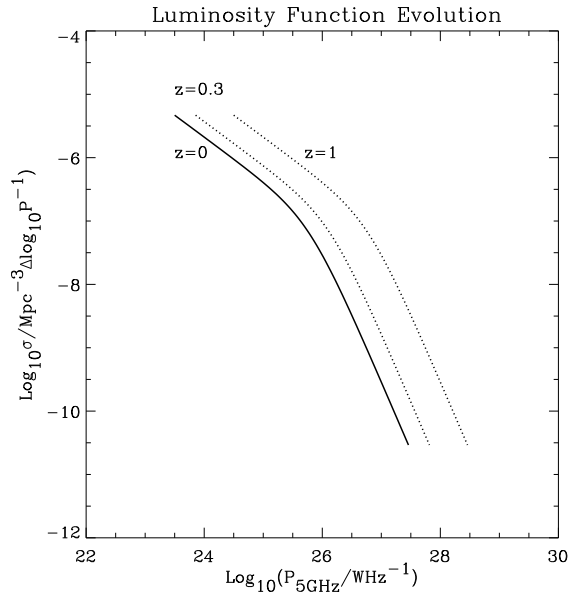,width=8cm}
\caption{\label{lfevolve} The evolution of the luminosity function of 
steep spectrum sources as determined by Dunlop and Peacock (1990).}
\end{figure}

The LLF for powerful radio sources and its cosmological evolution, has 
been studied by Dunlop \& Peacock (1990). We will use the pure luminosity
evolution model, since it fits the available redshift and source-count data 
well, and it is relatively straightforward to implement.
In this particular model, the overall shape of the 
luminosity function does not change with 
cosmological  epoch, only the normalisation in luminosity (see fig. \ref{lfevolve}).
Dunlop and Peacock (1990) parameterise an evolving two-power-law luminosity
function as
\begin{equation}
\rho(P_\nu,z)=\rho_o \left\{ \left( \frac{P}{P_c(z)}\right)^a + 
\left( \frac{P}{P_c(z)}\right)^b \right\}^{-1}
\end{equation}
where $a$ and $b$ are the two power-law slopes, $P_c(z)$ is the 
evolving `break' luminosity, and $\rho_o$ is determined by normalisation
at z=0. The redshift dependence, $P_c(z)$, was parameterised by 
Dunlop \& Peacock as 
\begin{equation}\label{eqcor}
\log P_c(z) = a_0 +a_1z+a_2z^2
\end{equation}
The best-fit model parameters for pure luminosity evolution ($\Omega_0=1$)
are, $\rho_o=-6.91$, $a=0.69$, $b=2.17$, $a_0=25.99$ (in W/Hz), 
$a_1=1.26$, $a_2=-0.26$.
Since Dunlop \& Peacock (1990) did their analysis at 2.7 GHz, their radio
powers have to be transformed to 5 GHz. Assuming a mean spectral index of 
$-0.75$, we use a conversion factor of $-0.20$ in the logarithm.
This luminosity evolution parameterisation, as shown in figure \ref{lfevolve},
 is used to derive the LLF of GPS sources.
First the radio powers, as given in table 1 and 
2, are corrected for the cosmological evolution of the luminosity 
function. This correction factor as function of redshift is equation \ref{eqcor}.
For example, at z=1, the luminosity function has shifted a factor 10
towards higher luminosities, and therefore 2128+048 with a radio power of 
$10^{27.8}$ W/Hz will contribute to the LLF at $10^{26.8}$ W/Hz.
Note that this correction is independent of luminosity, and therefore 
the difference in radio luminosity of young and old sources does not 
have to be accounted for. Note however, that the increase in number density
{\it is} dependent on radio luminosity due to a change in the slope of the 
luminosity function. The number densities increase from z=0 to z=1 by a 
factor of 5 and 150 for low and high luminosity sources respectively.

\begin{figure}
\psfig{figure=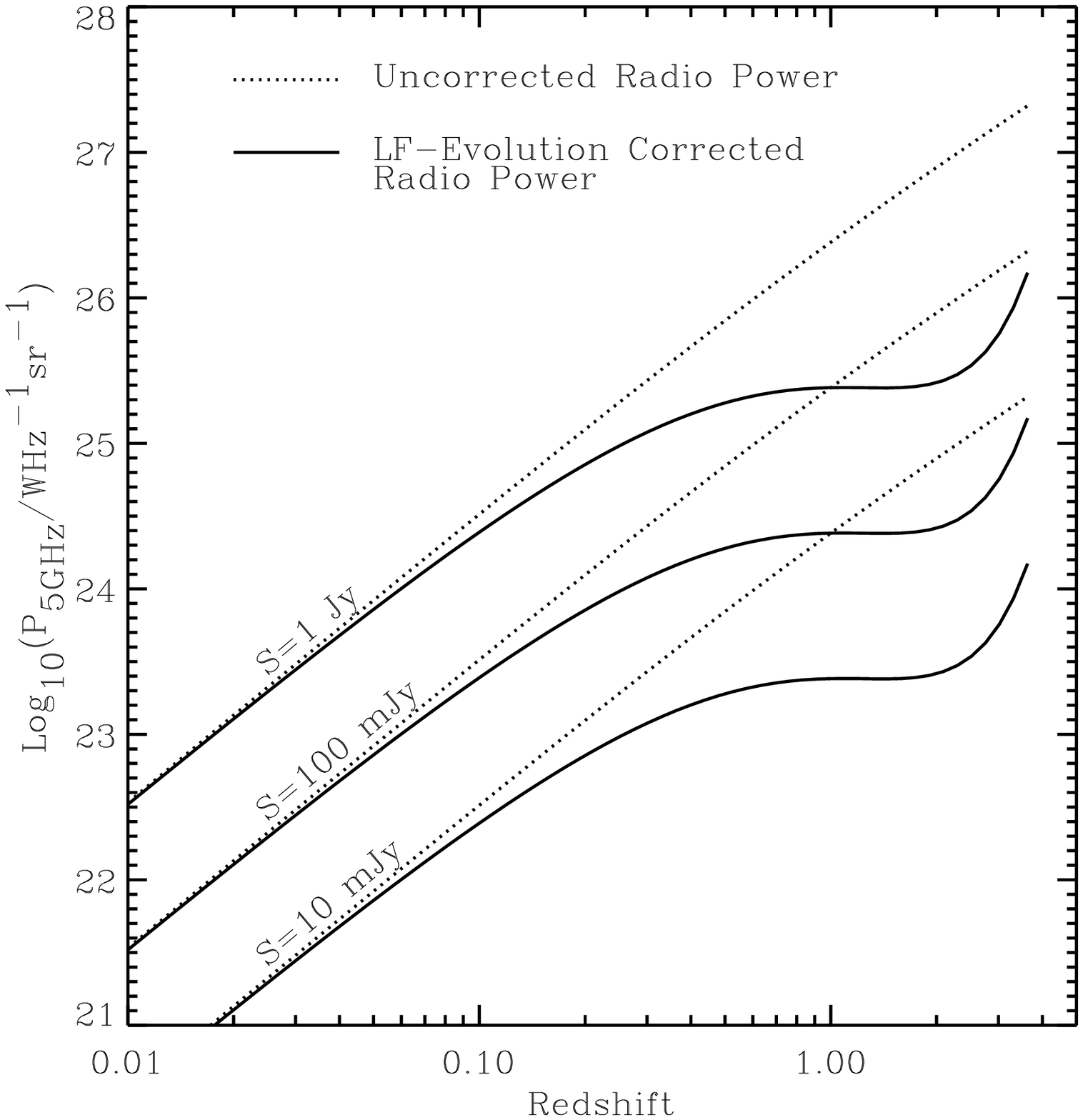,width=8cm}
\caption{\label{lumcor} The radio power for a source of 10 mJy, 100 mJy, and 
1 Jy as function of redshift, and its radio power corrected for the evolution
of the luminosity function, assuming $H_0$=50 km sec$^{-1}$Mpc$^{-1}$, and 
$\Omega_0$=1.}
\end{figure}

Figure \ref{lumcor} shows the corrected and uncorrected radio powers
for a source with flux densities of 10 mJy, 100 mJy and 1 Jy (assuming 
a spectral index of $-0.5$ at about 5 GHz).
Interestingly, the corrected radio power for a source with an certain observed
flux density, does not significantly change at $0.6<z<2.0$. Hence, although for
many GPS galaxies no spectroscopic redshift has been measured, this is 
not likely to influence the result, since they will probably all be 
in this redshift range.

The next step is to correct the number of sources observed for the volume 
of space over which they can be observed. As can be seen from figure 
\ref{lumcor}, a flux density limit in the sample of 1 Jy means that all
sources with corrected luminosities greater than 25.4 $W Hz^{-1}Sr^{-1}$
can be detected out to z=2, and that only source with lower radio power,
and consequently at $z<0.6$, have to be corrected for the fact that they
only could have been seen out to a certain redshift. However, 
a possible additional redshift limit results from the lower limit in peak 
frequency at 0.4 GHz in the bright Stangellini et al. sample, and 
in the faint sample due to the limit in 325-5000 MHz spectral index.
For these sources a weight-factor is used equal to the volume of the survey 
(assuming a redshift limit of 2.0) divided by the maximum volume over which 
they could have been in the sample, which is dependent on the maximum 
observable redshift. 
The corrections above are relatively straight-forward. However, some
additional, more complicated corrections have to be made for the 
faint GPS sample. Firstly, this sample is originally selected at 
325 MHz frequency, eg. on the optically thick part of their spectrum. 
Furthermore, only sources with positive spectral indices between
this frequency and 5 GHz were initially selected. Therefore the faint 
WENSS sample is more biased towards GPS sources with higher 
peak frequencies than the bright Stanghellini et al. sample. To correct
for this we assumed that the parent distribution of peak frequencies is 
independent of flux density and radio power, and determined what fraction
of the Stangellini et al sample would have been included in the sample
if it would have been selected as for the faint WENSS sample. 
It turns out that 26 \% of the galaxies in the bright sample have 325 MHz 
flux densities $>$ 1 Jy and positive spectral indices between 325 MHz and 
5 GHz. In addition, the bright GPS sample has a limit in optically thin
spectral index of $-0.5$, while several sources in the faint sample have 
a flatter optically thin spectral index. Taking into account these two 
effects, the number densities for the faint sample are multiplied by a factor 
3.2. 
\begin{figure}
\psfig{figure=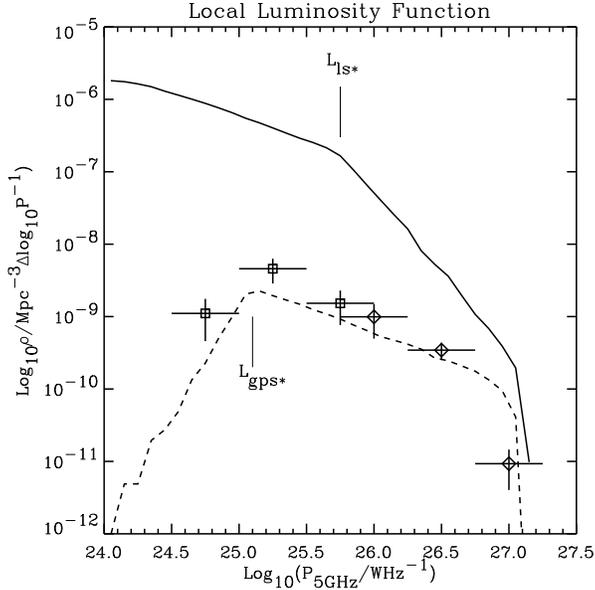,width=8cm}
\caption{\label{llf} The local luminosity function of GPS radio sources,
as derived from the bright sample of Stanghellini et al. (diamonds) and 
our faint GPS sample (squares). The dashed and solid lines give the 
simulated LLFs for GPS and large size radio sources respectively.
The model-parameters are chosen in such way that the simulated LLF of 
large size radio sources matches that of steep spectrum sources
as derived by Dunlop \& Peacock (1990).}

\end{figure}

The resulting local luminosity function of GPS sources is shown in 
figure \ref{llf}. Note that a luminosity bin (centred at 23.75 Watt/Hz) 
containing only a single source (B0830+5813) is omitted due to its 
large uncertainty.

We compared the resulting LLF with an LLF of a simulated radio source 
population of $10^6$ objects, with random ages, and a jet-power 
distribution defined as in equation 10.
The `observed' luminosity of a source was calculated assuming that it had evolved over its lifetime according to the 
luminosity evolution derived in section 4.1, out to a maximum size, $r_+$.
At $r<r_*$, the size of the source evolves as $r=t^{1/2}P_J^{1/4}$.
To avoid a discontinuity in propagation velocity 
at $r_*$, the source evolves from $r_*$ as,
\begin{equation}
r(P_J,t)=\frac{\gamma}{2}\gamma^{-\frac{1}{\gamma}}(\gamma P_J^\frac{1}{2}t)
^\frac{1}{\gamma}+(1-\frac{\gamma}{2})
\end{equation}
with $\gamma=(4-\beta)/2$. The luminosity of a source increases at $r<r_*$
as $L=P_J^{7/8}(r/r_*)^{2/3}$, and as $L=P_J^{7/8}(r/r_*)^{\frac{2}{3}-\frac{7}{6}\beta}$ at $r>r_*$. This results in a similar evolution for 
large size radio sources as derived in section 4.1, with the luminosity
at $r=r_*$ only dependent on $P_J$.
It was not our aim to determine absolute
values for number densities and radio powers with these simulations.
The results of the simulation were
scaled in such way, that the LLF obtained for large size radio sources,
matched the LLF of steep spectrum sources as derived by Dunlop \& peacock 
(1990). 

\begin{table}
\caption{\label{param} Dependence of LLF characteristics on 
 the model parameters}
\begin{tabular}{ccr}
Characteristic of LLF   &                 Dependence& Value used \\
          \\
LS-LLF slope at high $L$&$\frac{5+7\beta}{4-7\beta}$& $-$3.17$^a$\\
\\
LS-LLF slope at low  $L$&$(\delta+\frac{1}{2})\frac{1}{\kappa}-1$&$-$1.69$^a$\\
\\
$L_{max}$ & $ P_+^{\kappa}$ & $10^{27.1}$ \\
\\
$L_{ls*}$   & $P_+^{\kappa}\left(\frac{r_{+}}{r_{*}}\right)^
{\frac{2}{3}-\frac{7}{6}\beta}$ & $10^{25.8^a}$ \\
\\
$L_{gps*}$ & $P_{-}^{\kappa}$ & $10^{25.0}$ \\
\\
\end{tabular}
\begin{tabular}{l}
$^a$ Value chosen to match the LLF of Dunlop \& Peacock (1990).
\end{tabular}
\end{table}

Table \ref{param} lists the important characteristics of the 
simulated LLF of large size and GPS sources, and their dependence on
the free model parameters. 
The parameters $\delta$ and $\beta$, as defined in equations 10 and 3, 
determine the slope of the low and high luminosity part of the 
LLF of large size radio sources. These were chosen to be similar 
to the parameters $a-1$ and $b-1$ as derived by Dunlop \& Peacock (1990),
with $\delta=-1.10$ and $\beta=1.16$.
This value of $\beta$ is slightly lower than derived from 
X-ray observations of nearby ellipticals ($\beta=1.5-2$, 
Trinchieri et al. 1986). 
Note however, that the radio source population is dominated by 
objects with size $>20$ kpc, for which the surrounding medium is 
dominated by intra-cluster gas, which is expected to have a flatter
density gradient.
With the parameters $\delta$ and $\beta$ and the slopes 
of the LLF of large size radio sources fixed, the relative positions of 
the break luminosities could be determined. A sharp cut-off will occur
near the highest luminosity, $L_{max}$.
The number of GPS galaxies in the highest luminosity bin, as 
shown in figure \ref{llf}, is lower than expected from the extrapolation 
of the LLF at lower luminosities. This can be explained if this luminosity
bin is near the cut-off luminosity $L_{max}$. We therefore chose log $L_{max}$ to be 27.1 (W Hz$^-1$). The break luminosity of large size 
radio sources, is also determined by Dunlop \& Peacock (1990)
to be log $L_{LS*}$ = 25.79 (corrected to 5 GHz).
As can be seen from table \ref{param}, the luminosity ratio
$L_{max}/L_{ls*}$ determines the value of $r_{+}/r_{*}$. 
 This corresponds to 
a maximum size for a radio source of 100 kpc, assuming $r_*=1$ kpc. 
This value is quite near the turnover seen in the linear size
distribution of 3CR galaxies, as shown by O'Dea \& Baum (1997).
The break luminosity of GPS sources, $L_{gps*}$, relative to $L_{max}$,
is dependent on the range of jet-powers $(P_+/P_-)$.
To let $L_{gps*}$ coincide with the peak in the observed GPS LLF, 
a value of $(P_+/P_-)$=200 was used.

Although the uncertainties on the datapoints are large and several free
parameters enter the simulation, 
figure \ref{llf} shows that the shape of the LLF of GPS 
sources is as expected. Note that most free parameters are determined
by fitting the LLF of large size radio sources to that of Dunlop \&
Peacock (1990), except $r_{+}$ and $(P_+/P_-)$.
This analysis should be
regarded as an example of how future large and homogeneously
defined samples of GPS sources can constrain the luminosity
evolution of extragalactic radio sources.

The proposed increase in luminosity for young radio sources seems to 
be in contradiction to the high number counts of GPS sources with respect
to large size radio sources suggesting that they should decrease in 
radio luminosity by a factor $~10$  during their lifetime (Fanti et al. 1995,
Readhead et al. 1996, O'Dea \& Baum 1997). However, this is not the case.
Flux density limited samples, as used for these analyses, only 
probe the most luminous objects at any redshift. As can be seen
from figure \ref{llf}, at high luminosities, the two luminosity function
approach each other, due to the flatter slope of the 
luminosity function of GPS sources.
This results in a relatively high number density of GPS sources in
flux density limited samples. 

\subsection{Summary and Conclusions}

In this paper we show that in addition to the well known correlation
between spectral peak frequency and angular size (eg. Fanti et al. 1990),
a correlation exists between the peak flux density and angular size
of GPS \& CSS sources. The strength and sign of these correlations 
are exactly as expected from SSA theory, assuming equipartition, and are 
therefore a strong indication that SSA is indeed the cause of the spectral
turnovers in these objects. Furthermore, these correlations are consistent
with GPS \& CSS sources evolving in a self-similar way.
Interestingly, the self-similar evolution scenario is 
better fitted by assuming an equipartition than a constant magnetic field.

In flux density limited samples, GPS galaxies are found at higher 
redshifts than large size radio sources. Since 
the lifetimes of radio sources are short compared to cosmological timescales,
this can only mean that the slope of their luminosity functions
are different, if GPS sources are to evolve into large size radio sources.
It is shown that that the slope of a luminosity function is strongly
dependent on the evolution of radio power of the individual sources. 
A new method is introduced to constrain the luminosity evolution 
of radio sources using the luminosity functions of `young' and `old' 
objects. It is shown that if GPS sources are increasing in radio power 
with time, it would result in a relatively flatter slope of their luminosity
function compared to that of large size radio sources which decrease in 
radio power.

A simple model was developed in which a radio source, embedded in 
a King profile medium, evolves in a self similar way under the 
equipartition energy assumption.
 This model indeed results in 
the suggested increase in luminosity for young radio sources, 
and decrease in luminosity for old, extended objects.
The calculated luminosity function for large size radio sources shows  
a break and slopes at low and high luminosity comparable to that derived by
Dunlop \& Peacock (1990) for steep spectrum sources. 
The local luminosity function (LLF) of GPS sources can not be
measured directly since so few GPS sources are found at low redshift.
Therefore, the knowledge of the cosmological evolution of the luminosity
function of steep spectrum sources, as derived by Dunlop \& Peacock (1990),
is used, so that the LLF can be derived from the complete samples
of bright and faint GPS sources. It is shown that this LLF is as expected for
radio sources which increase in luminosity with time, which is confirmed by 
simulations of the young radio source population. 
Note however, that the bright and faint GPS samples are constructed in 
very different ways and that therefore large corrections had to be made. 
Uncertainties are still very large and ideally large samples of GPS sources
should be constructed which are uniformly selected at low and high flux 
densities.

\section{Acknowledgements}

{\it We like to thank the anonymous referee for carefully reading the 
manuscript and valuable suggestions.
We also thank Christian Kaiser and Jane Dennett-Thorpe for helpful comments.
This research was in part funded by the European Commission under
contracts  ERBFMRX-CT96-0034 (CERES) and 
ERBFMRX-CT96-086 (Formation and Evolution of Galaxies), and
SCI*-CT91-0718 (The Most Distant Galaxies).}

\end{document}